\documentclass[%
 aip,
rsi,%
 amsmath,amssymb,
preprint,%
]{revtex4-1}

\usepackage{graphicx}% Include figure files
\usepackage{dcolumn}% Align table columns on decimal point
\usepackage{bm}% bold math
\usepackage{subfigure}
\usepackage{color}
\usepackage[normalem]{ulem}
\begin{document}

\preprint{AIP/123-QED}

\title{The Nanofluidic Confinement Apparatus: Studying confinement dependent nanoparticle behavior and diffusion}% Force line breaks with \\

\author{Stefan Fringes}
\altaffiliation[Also at ]{Department of Chemistry, University of Zurich.}%Lines break automatically or can be forced with \\
\affiliation{ 
	IBM Research - Zurich, S\"{a}umerstr. 4, 8803 R\"{u}schlikon, Switzerland
	%\\This line break forced with \textbackslash\textbackslash
}

\author{Felix Holzner}
\altaffiliation[Present address: ]{SwissLitho AG, Technoparkstrasse 1, 8005 Zurich.}%Lines break automatically or can be forced with \\
\affiliation{ 
	IBM Research - Zurich, S\"{a}umerstr. 4, 8803 R\"{u}schlikon, Switzerland
	%\\This line break forced with \textbackslash\textbackslash
}
\author{Armin W. Knoll}%
 \email{ark@zurich.ibm.com.}
\affiliation{ 
IBM Research - Zurich, S\"{a}umerstr. 4, 8803 R\"{u}schlikon, Switzerland
%\\This line break forced with \textbackslash\textbackslash
}%

\date{\today}% It is always \today, today,

\begin{abstract}	
We present a versatile setup for investigating the nanofluidic behavior of nanoparticles as a function of the gap distance between two confining surfaces. The setup is designed as an open system which operates with small amounts of dispersion of $\approx 20\,\mu$l, permits the use of coated and patterned samples, and allows high-numerical-aperture microscopy access. Piezo elements enable 5D relative positioning of the surfaces. We achieve a parallelization of less than $1\,$nm vertical deviation over a lateral distance of $10\,\mu$m. The vertical separation is tunable and detectable with subnanometer accuracy down to direct contact. At rest, the gap distance is stable on a nanometer level. Using the tool we measure the vertical position termed height and the lateral diffusion of $60\,$nm charged Au nanospheres as a function of confinement between a glass and a polymer surface. Interferometric scattering detection results in sub $10\,$nm vertical and sub $5\,$nm lateral particle localization accuracy, and a single particle illumination time below $40\,\mu$s. 
%The short illumination time corresponds to a bulk diffusion length of $\approx 17\,$nm which is needed to provide a good localization of the particles. 
We measure the height of the particles to be consistently above the gap center, corresponding to a higher charge on the polymer substrate. In terms of diffusion, we find a strong monotonic decay of the diffusion constant with decreasing gap distance. This result cannot be explained by hydrodynamic effects, including the asymmetric vertical position of the particles in the gap. Instead we attribute it to an electroviscous effect. For strong confinement of less than $120\,$nm gap distance, we detect an onset of sub-diffusion which can be correlated to a motion of the particles along high-gap-distance paths.
%Using the pixel by pixel interference phase signal we determine \emph{in situ} the optical path difference in the gap which is closely related to the confinement gap modulation. Thus we can establish a correlation between the local gap distance and the amount of sub-diffusion. The results corroborate the interpretation of sub-diffusion as a motion of the particles along high gap distance trajectories. 
%The nanofluidic confinement apparatus can even resolve the influence of the subnanometer surface roughness, which decreases the diffusion of the particles even further. The strong influence of surface topography on the particle movement in this tunable confinement offers the opportunity to use the nanofluidic confinement apparatus to deliberately assemble or sort nano-objects. 
%in situ tuning of the nanofluidic confinement with nanometer accuracy in combination with simultaneous fast 3D tracking of the nano-objects enables numerous novel opportunities for studies and deliberate manipulation of a broad choice of nano-objects. 
\end{abstract}

%Use showkeys class option if keyword
\maketitle

\section{\label{sec:intro}Introduction}
A fundamental understanding of the motion of micro- and nano-scaled objects in nanofluidic confinement is important for many biological and technical processes such as the anomalous diffusion in cellular environments,\cite{Regner13bpj,Baum2014} the delivery of drugs,\cite{Langer03AIChE} the formation of colloidal crystals,\cite{Gong02la,Reinmueller12jcp} particle sorting,\cite{Huang04sci} and directed self-assembly.\cite{Grzelczak2010}

Nanofluidic systems in general are characterized by spatial distances in at least one dimension of less than 100 nm. This distance range interferes with several natural length scales of particle-surface interactions \cite{Bocquet2014}, such as the electrostatic interactions. The electrostatic interactions between charged objects and surfaces in a nanofluidic system decay approximately exponentially with separation and a characteristic length scale termed Debye length.\cite{Hunter87colloid} 
%Because of this strong distance dependance the behavior of nanofluidic systems is also extremely sensitive to the degree of confinement as measured by the surface distances. 
Experimentally, the gap-distance-dependent forces between two curved surfaces were studied in micro-rheology experiments \cite{dhinojwala1997micron,clasen2004gap} and in detail using the surface force apparatus~\cite{Israelachvili92book}.
%, and the wealth of results obtained is impressive. 
However, so far, most nanofluidic experiments involving confined particles have been performed using static surfaces and fixed geometries, which do not allow the degree of confinement to be varied \textit{in situ}. 

Recently it was demonstrated that the gap-distance-dependent electrostatic forces can be exploited to achieve geometry-induced trapping and manipulation of charged nanoparticles and vesicles in nanofluidic systems.\cite{Krishnan10nature} In a follow-up experiment, it was shown that crucial information on the trapping potential can be gained by using an AFM-type system and a micro-capillary to adjust the gap distance.\cite{TaeKim14natcom} 

Another example of a strongly gap-dependent behavior is the lateral diffusion of particles in a nanofludic gap. In microfluidic systems, it has been shown that the theoretical predictions of hydrodynamically hindered diffusion are in agreement with the measured diffusivity of microparticles.\cite{Lin00PhysRevE,Dufresne01epl} However, in nanofluidic systems, a $50 - 70\,\%$ lower diffusion is observed when geometrical dimensions approach the Debye screening length $\cite{Kaji06anabiochem,Eichmann08lgm,Zhao16anachem}$. The mechanisms that have been proposed to explain the increased hindrance are anomalous viscosity\cite{Kaji06anabiochem}, anomalous diffusion\cite{Zhao16anachem} and an electroviscous effect.\cite{Eichmann08lgm} 
%Although these effects depend on the confinement, measurements were so far performed only at a fixed gap distance. %no gap distance dependent behavior is analyzed for a particular particle ensemble.

Here we present a versatile setup that allows the distance between two parallel confining surfaces for samples of choice and a cover-glass to be adjusted and measured with nanometer accuracy. First, we describe and characterize the system, 
%An interferometric scattering detection (iSCAT) [citation] method is used to analyze the three-dimensional (3D) motion of nano-objects. 
%The method yields a good imaging contrast due to the enhanced coupling of light to the surface plasmons at the resonance frequency and due to the interference of light scattered by the particles with light reflected by the sample and the cover-glass. The interference signal away from the particle positions is used to determine the separation of the confining surfaces with sub nanometer precision. 
and then demonstrate its utility by measuring the behavior of $60\,$nm charged Au nanospheres in confinement between a glass and a polymer surface. We first determine the height of the particles as a function of gap distance by means of their varying optical contrast. Next we determine the lateral diffusion for a range of fixed gap distances. The gap-dependent measurement allows us not only to measure the decreasing diffusion coefficients but also to determine the onset of a scale dependent diffusion induced by the roughness of the confining surfaces. A comparison with theory indicates that hydrodynamic effects alone cannot explain the behavior observed. 
\section{\label{sec:method}Method}

\subsection{\label{sec:setup}Nanofluidic confinement apparatus}
\begin{figure}
	\centering
	\includegraphics[width=.7\linewidth]{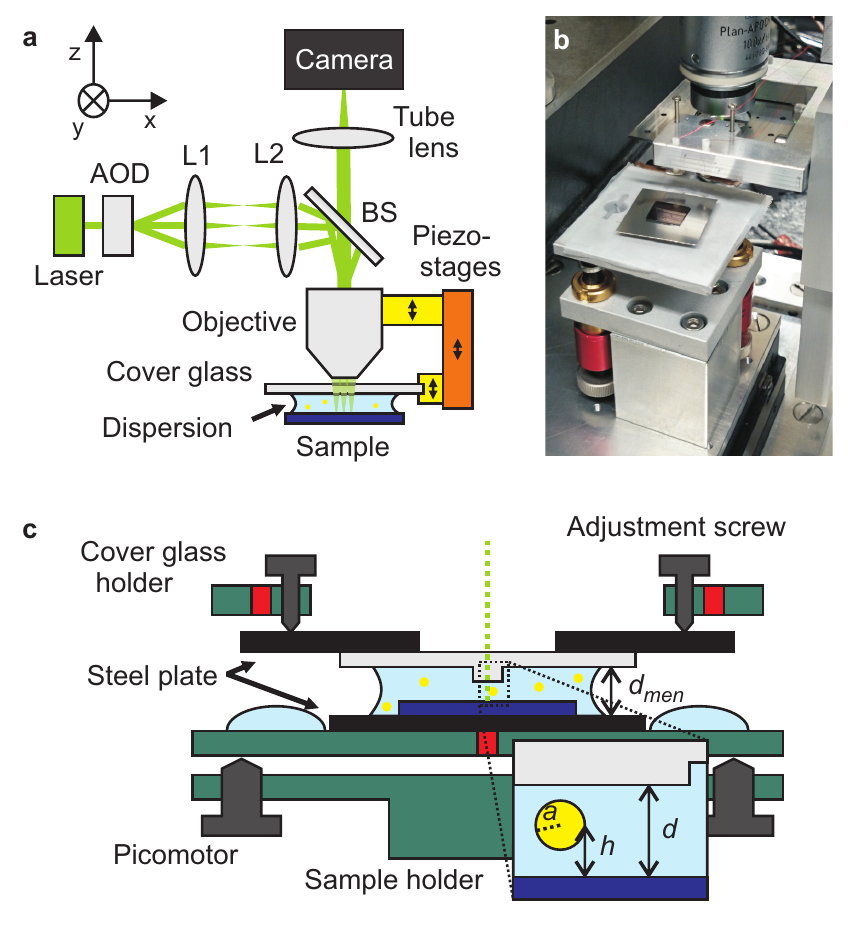}
	\caption{(a) The optical setup consist of a laser, an acousto-optic deflector (AOD), and a telecentric system (L1 and L2) for scanned laser illumination of the sample through a beam splitter (BS), and an oil immersion objective. Two linear piezo-stages (yellow) and one coarse-positioning stage (orange) allow to fine tune the focus and the confinement independently and to access a large range of gap distances. (b) Photograph of the nanofluidic confinement apparatus. (c) Sketch of the vertical profile of the system, see text for details. The inset visualizes the nanofludic slit with gap distance $d$ confining a particle with radius $a$ at height $h$.}
	\label{fig:setup}
\end{figure}

A schematic illustration of the nanofluidic confinement apparatus is shown in Fig. \ref{fig:setup}(a). The optical illumination and detection scheme is based on interferometric scatterning detection (iSCAT) and was described in detail elsewhere\cite{Jacobsen06optexp,Kukura09natmet,Mojarad13optexp,Fringes2016jap}, here we just provide a brief description.

By raster scanning the focus of a 532$\,$nm continuous-wave laser (Samba $50\,$mW, Cobolt), the sample area of interest is illuminated. Scanning and focusing are done by a two-axis acousto-optic deflector (DTSXY, AA Opto-Electronic), a telecentric system, and a 100$\times$, 1.4 numerical aperture (NA) oil-immersion objective (Alpha Plan-Apochromat, Zeiss). The reflected light is collected by the same objective, and images are captured by a high-frame-rate camera (MV-D1024-160-CL-12, Photon Focus). Typically we use a field of view of $300 \times 300$ pixels, corresponding to an area of $33 \times 33\,\mu$m$^2$. The imaging rate is 800 frames per second (FPS), given by the exposure time of $0.75\,$ms and a trigger delay of $0.5\,$ms, which was selected to avoid frame drops. We achieve uniform illumination using a single scan per frame and a laser line spacing of $500\,$nm, which is consistent with an estimated laser spot size of $\approx\,2\,\mu$m. Accordingly, a single point on the sample is scanned by $\le 4$ laser lines of  $10\,\mu$s duration corresponding to a total time of $\tau_{\mathrm{avg}}\lesssim\,40\,\mu$s. During $\tau_{\mathrm{avg}}$ the diffusion of a $60\,$nm Au nanoparticle (bulk diffusivity $D_p \approx 7 \mu$m$^2$s$^{-1}$) in one dimension is $\lesssim \sqrt{2 D_p \tau_{\mathrm{avg}}} \approx 25\,$nm, which is small compared to the laser line spacing. Thus the image taken by the camera contains information about the position of the particle averaged over a duration of $\lesssim\tau_{\mathrm{avg}}$.

The mechanical part with the tunable confinement setup is mounted below the objective (see Fig. \ref{fig:setup}\,(b)). A schematic cross section through the center of the system is sketched in Fig. \ref{fig:setup}\,(c) (not to scale): A droplet of particle dispersion is confined by the cover-glass (light gray) and the sample (dark blue). The glass and the sample are both glued to steel plates (black). Magnets (red) in the aluminum holders (green) fix the position of the steel plates. Three adjustment screws are used to align the tilt of the cover-glass with respect to the focal plane of the objective. Parallelization of the substrate to the cover-glass is done by three linear piezo actuators (Picomotor, Newport). The distance of the cover-glass and the microscope objective relative to the substrate is controlled by two linear piezo-stages (100$\,\mu$m, Nano-OP100, Mad City Labs), which are attached to a coarse-positioning stage (MT-84, Feinmess). A mesa is etched in the cover-glass such that the area outside the mesa is recessed by $\approx 50\,\mu$m (see next section for details). The mesa provides good optical access to the nanofluidic region and ensures that the gap distance \textit{d} between the cover-glass and sample (see inset of Fig. \ref{fig:setup}\,(c)) can be reduced until a colloid has intimate contact to both surfaces.

A droplet volume of $V_{\mathrm{drop}}\ge20\,\mu$l is required such that the dispersion overflows the sample and wets the metal holder. This geometry increases the distance at the meniscus $d_{\mathrm{men}}$ to approximately $600\,\mu$m (sample thickness $550\,\mu$m). Therefore also the radius of curvature of the droplet is increased, resulting in a reduced Young--Laplace pressure and a high stability of the system. A water reservoir next to the the central droplet (Fig. \ref{fig:setup}\,(c)) reduces the evaporation of the droplet in the slit and ensures system stability for several hours.

\subsection{Cover-glass and sample preparation}
The mesa of the cover-glass (D263T borosilicate, UQG) was fabricated as follows: First, a masking layer of $30\,$nm Cr and $300\,$nm Au was sputtered onto the glass. Second, a photoresist (AZ4533, MicroChemicals) was spin coated and patterned by photolithography. Third, the masking layer was removed by wet etching (TechniEtch ACI2, MicroChemicals and TechniStrip Cr01, MicroChemicals) of the unprotected areas, leaving behind a central metal-resist stack defining the position of the mesa. The area around the stack was etched for $75\,$s by concentrated hydrofluoric acid  ($49\,\%$ HF) to define the mesa. A mesa height of 40--45$\,\mu$m was measured with a profilometer (Dektak, Veeco), corresponding to an etch rate of $\approx 36\,\mu$m/min, similar to the rate observed by Zhu et al.\cite{Zhu09D263etching} Finally, the remaining masking layer was removed by etching, and the processed cover-glass was cleaned by peeling off a polymer layer (Red First Contact, Photonic Cleaning Technologies), in a helium plasma (Piezobrush, Relyon Plasma) for $20\,$s and by rinsing with ultrapure water (Millipore, $18\,\mathrm{M}\Omega\mathrm{cm}$).

A $52\,$nm thick cross-linking polymer (HM8006, JSR) was spin coated onto a silicon sample to increase adhesion for the subsequently spin coated $175\,$nm thick poly-phthalaldehyde (PPA) film. The thicknesses were measured with AFM. The refractive indices $n_{HM} = 1.67$ and $n_{PPA} = 1.59$ were measured by ellipsometry.\\
A colloid of citrate stabilized $60\,$nm Au nanospheres (BBI Solutions) with a manufacturer-specified diameter of $2a=59.8\pm4.8$ and density of $\approx 2.6\times10^{12}$ particles per ml was diluted 1:10 in fresh ultrapure water (Millipore, $18\,\mathrm{M}\Omega\mathrm{cm}$) to reduce the ion concentration. The diluted dispersion was used within a few hours. A pH of $6.8\pm0.2$, a zeta potential of $\zeta = -58\,$mV, a specific conductivity of $\Lambda = 11.5\,\mu\mathrm{Scm}^{-1}$, and hydrodynamic diameter of $2a=62.1\,$nm
% $2a=62.1\pm0.2$
were measured for a 1:150 diluted dispersion using a Malvern Zetasizer. 
We observed a linear dependency between the conductivity and the degree of dilution, which is expected for strong electrolytes such as sodium citrate and sodium chloride. Both can be present, since the synthesis involves the reduction of chloroauric acid (HAuCl$_4$) by sodium citrate (Na$_3$Cit).\cite{Frens73controlled} The citrate also functions as a capping agent, therefore we first determine the cation concentration from the conductivity measurement and then estimate an upper limit for the Debye length of $\kappa^{-1}\approx 8.9$\,nm for the 1:10 diluted colloid. In an independent measurement, we determined a larger Debye length for the same but more diluted colloid, consistent with the Debye length presented here.\cite{Fringes2016jap}

%\subsection{Performance}
\subsection{Measurement of gap distance and stability of the mechanical setup}

The performance of the setup is characterized by the precision achieved in controlling and detecting the gap distance. For a slit filled with aqueous dispersion, a change in gap distance leads to a change in the Young--Laplace pressure, which bends the cover-glass such that the motion of the piezo and the cover-glass are not in 1:1 correspondence. Therefore we use the interference of the light between the sample and the cover-glass as a measurement of the gap distance.

\begin{figure}
	\centering
	\includegraphics[width=.7\linewidth]{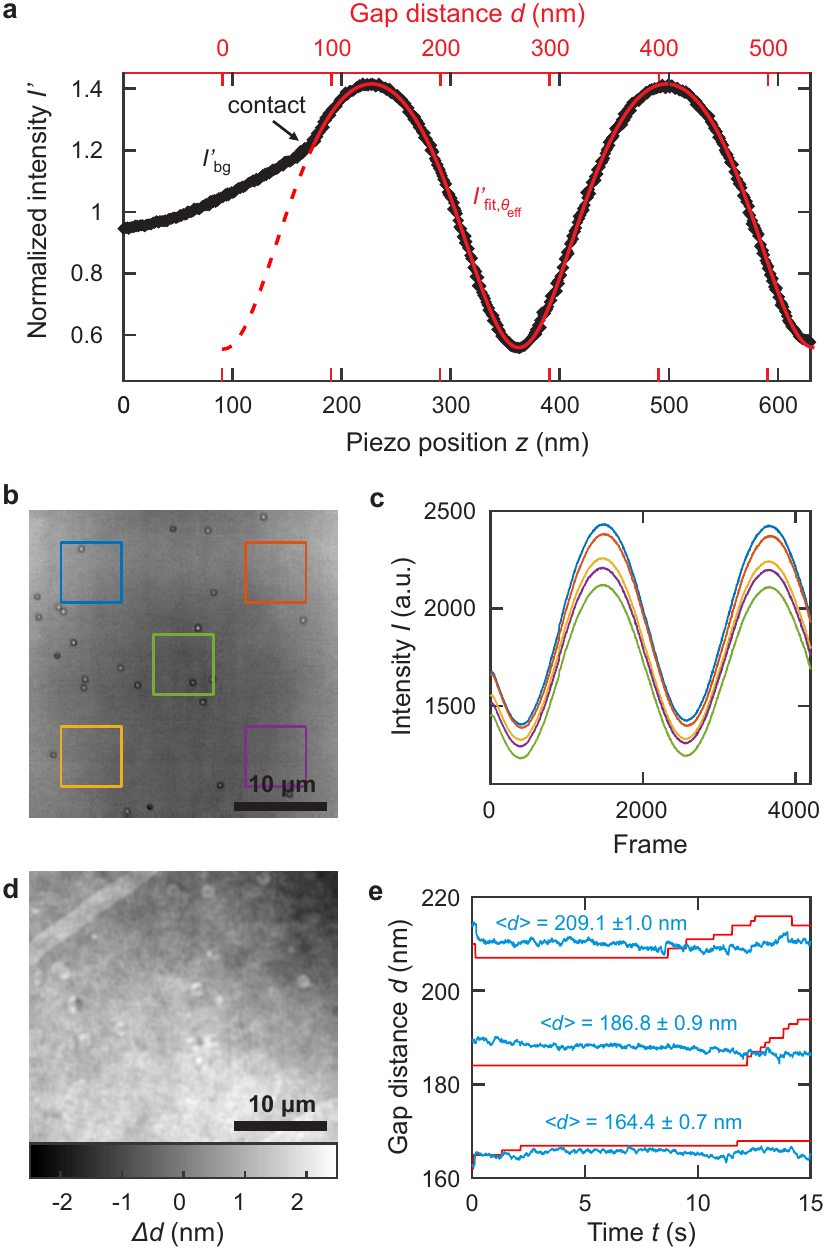}
	\caption{(a) Measured intensity reflected from a glass-air-silicon slit (black) while the cover-glass is displaced vertically with a piezo motor. The red line depicts the result of a fit to our optical model. (b) Typical raw image of $60\,$nm Au nanospheres in the nanofluidc slit. (c) The median intensity captured in the areas indicated by the boxes in (b), while increasing the gap distance by $1\,$nm every 10th frame. (d) Effective gap distance variation $\Delta d$ in the nanofluidic slit obtained from the local variation in optical path difference. (e) The height of the cover-glass (red) is adjusted by a feedback loop to ensure a constant gap distance (blue) during experiments.}
	\label{fig:performance}
\end{figure}

For this measurement, we have to consider light rays departing from normal incidence, because we use a high NA objective to focus and collect the light. We address this issue by determining an effective incident angle as described in detail in Ref.\cite{Fringes2016jap}. The angle is determined from a measurement of the normalized interference intensity $I'$ as a function of the cover glass position $z$ in air to avoid the effect of the pressure changes mentioned above, see Fig. \ref{fig:performance}\,(a). The signal arises from the interference of light rays reflected by the interfaces of the glass-water-polymer-silicon stack. We have developed an optical model\cite{Fringes2016jap} based on the transfer-matrix method, that considers the focusing of a Gaussian laser-beam. The result of a fit to the data is shown as red dashed line in Fig. \ref{fig:performance}\,(a). Fit parameters are the effective incident angle $\Theta_{\mathrm{eff}}=5.9^\circ$ and the phase of the signal. The phase of the signal and the first contact point at a gap distance of $d \approx 80\,$nm fixes the absolute gap distance (see red axis). The required refractive indices for silicon, $n_{\mathrm{Si}}=4.14$, and for the cover-glass, $n_{\mathrm{D263}}=1.53$, are taken from literature. To measure the gap distance in the water-filled system, we use the optical model and propagate the effective incident angle into the dielectric layers by using Snell’s law.
%This angle considers the focusing of light and is relatively small due to the strong under filling of the objective. Second, we model the reflectivity of the plan-parallel multi-layer stack with the transfer-matrix method. Each layer is characterized by a reflection and propagation part. The required refractive indices for silicon $n_{Si}$=4.14, for water $n_{H_2O}=1.33$ and for the cover-glass $n_{D263}$=1.53 are taken from literature. The parameters of the polymer films are $t_{HM}=52\,$nm, $t_{PPA}=175\,$nm and $n_{HM} = 1.67$, $n_{PPA} = 1.59$ (see Sample preparation). Third, the raw images are normalized and filtered by their spatial and temporal neighborhood into background intensity and particle contrast. Distinct gap distances can be assigned by matching the measured and simulated background intensity values. The details of these steps are described elsewhere \cite{Fringes2016jap}.

Parallelization of the surfaces is achieved by measuring the interference signal in the four corners and at the center of the illuminated area (see Fig. \ref{fig:performance}\,(b)): From the relative phase shift of the respective signals (see Fig. \ref{fig:performance}\,(c)), the tilt of the confining surfaces can be determined. By tilting the sample, the phase difference was minimized using the cross-correlation of the corner to the center signals. 

The optical path difference between glass and substrate varies because of the inherent surface roughness of the contributing interfaces. This fact leads to a varying phase shift of the interference signal pixel by pixel. AFM measurements yield the following root-mean-square (RMS) roughnesses: $S_q^{\mathrm{D263}}\approx0.4\,$nm for the cover-glass, $S_q^{\mathrm{PPA}}\approx0.3\,$nm for the polymer surface and $S_q^{\mathrm{Si}}\approx0.2\,$nm for the silicon wafer. Since the silicon wafer is relatively flat and the refractive indices of polymer and glass are similar we approximate that all the phase differences originate from a roughness in the cover glass. The conversion from the phase shift to the gap distance is performed using the optical model mentioned above. The resulting gap distance image Fig. \ref{fig:performance}\,(d), reveals a remnant tilt between the two confining surfaces, which could be corrected further. Without this correction, we achieve a height difference of $3\,$nm over a distance of $30\,\mu$m. The standard deviation of the plane corrected gap distance image is $S_q^{\Delta{}d}\approx 0.6\,$nm, which is consistent with the measured surface roughness values.

During the measurements described in the subsequent sections, thermal drift and pressure changes may lead to a deflection of the relatively compliant cover-glass. These deflections are compensated by implementing a closed-loop system, that registers changes in the background interference intensity and adjusts the height of the cover-glass to keep the intensity constant. The feedback-loop can also operate during acquisition with a frequency of $20\,$Hz as illustrated by the red lines in Fig. \ref{fig:performance}\,(e)). The blue lines indicate the measured laterally averaged gap distances for $15\,$s.

%This duration is already sufficient for measuring the lateral confined diffusion, but the system has been tested to keep the confinement constant for several hours.

\subsection{Particle localization}

Radial symmetry-based tracking was used to identify the central lateral position of the nanosphere. This tracking algorithm yields similar  accuracies compared to Gaussian fitting, is fast in execution, and detects any radially symmetric intensity distribution.\cite{Parthasarathy12natmet} In particular the latter is important to detect the position at interference conditions for which the particle contrast vanishes at the center and only a diffraction ring of finite intensity is measured. We estimate an average lateral localization precision of $\le5\,$nm from the scatter of 35,000 detected positions obtained from 7 immobilized  particles. This precision is in agreement with simulated particles\cite{Parthasarathy12natmet} with a similar signal-to-noise ratio (SNR) of $\approx20$. We like to point out that we measure the same SNR using raw images similar to that in  Fig. \ref{fig:performance}\,(b), but for moving particles we can reduce the fixed-pattern camera pixel noise of the background by subtracting the temporal median of the image stack. With that correction, we obtain a SNR of $\approx 50$, which corresponds to a localization accuracy of less than $1.5\,$nm.\cite{Parthasarathy12natmet}

%Using this model the contrast of a sphere with radius $a$ can be simulated for all possible heights within the nanofluidic gap ($a \leq h \leq d-a$).

%\section{\label{sec:theory}Theory}
%\subsection{Potential energy profile in the slit}
%\begin{eqnarray}
%\Phi = \Phi_{PPA}e^{-(d-h-a)\kappa} + \Phi_{SiO_2}e^{-(h-a)\kappa}\\
%=> \log \left( \frac{\Phi_i} {\Phi_{ref}} \right) = -\frac{d_i-d_{ref}}{2}\kappa
%\end{eqnarray}

\section{Confined lateral diffusion}

In the following we first revisit briefly the existing hydrodynamic models describing confined lateral particle diffusion. According to these models, the diffusivity depends not only on the gap distance but also on the vertical position of the particles in the gap. To test these predictions, we included in our measurements described in the subsequent sections not only the diffusion but also the height of the particles in the gap.

\subsection{Hydrodynamic models}
Following the work of Eichmann \textit{et al.} \cite{Eichmann08lgm}, we present the linear superposition (LSA) and the coherent superposition approximation (CSA) to calculate the hindered lateral diffusion in a fluidic slit. A third approximation, the matched asymptotic expansion (MAE), is not considered here as it deviates only slightly from the LSA.

The diffusion coefficient of a freely moving spherical particle obeys the Stokes--Einstein-equation
\begin{equation}
D_0 = \frac{kT}{6\pi\eta a},
\end{equation} 
where $k$ is Boltzmann's constant, $T$ is the absolute temperature, and $\eta$ is the dynamic viscosity of the continuous medium. The hydrodynamically hindered diffusion parallel to a single interface is conveniently given by a correction factor $f_{||1}$: 
\begin{equation}
D_{||1}\left(h,a\right) = D_0\,f_{||1}(h,a).\\
\end{equation} 
Solutions are given in terms of the dimensionless particle height, $\omega=h/a$, for \cite{Pawar93IndEngChemRes}
\begin{align}
\omega>1.1:\nonumber\\
%	f_{||1}(h,a)&=1-\frac{9}{16}\omega^{-1}+\frac{1}{8}\omega^{-3}-\frac{45}{256}\omega^{-4}+\ensuremath{\mathcal{O}}(\omega^{-5})&\nonumber\\
f_{||1}(h,a)&=1-\frac{9}{16}\omega^{-1}+\frac{1}{8}\omega^{-3}-\frac{45}{256}\omega^{-4}-\frac{1}{16}\omega^{-5}&\nonumber\\
&+0.22206\omega^{-6}-0.205216\omega^{-7}&\\	
\omega\le1.1:\nonumber\\
	f_{||1}(h,a)&=1-\frac{15/8}{\ln(\omega-1)} &\nonumber\\
	&+e^{1.80359(\omega-1)}+0.319037(\omega-1)^{0.2592}&	
\end{align}
by Fax\`en \cite{Faxen23} and Goldman \cite{Goldman67chemeng}, respectively. A similar approach leads to the drag-reduced diffusion in a slit\cite{Oseen27hydrodyn}:
\begin{equation}
D_{||2}\left(h,a,d\right) = D_0\,f_{||2}(h,a,d),
\end{equation} 
where \textit{d} is the gap distance of the confining walls. Oseen suggested the LSA\cite{Oseen27hydrodyn}
\begin{equation}
	f_{||2}^{\mathrm{LSA}}(h,a,d) = \left[f_{||1}(h,a)^{-1}+f_{||1}(d-h,a)^{-1}-1\right]^{-1},
	\label{eq:LSA}
\end{equation}
where the drag of each wall is treated independently and the total force is given by the sum of the contributions.

Anoher expression, the CSA
\begin{align}	
	f_{||2}^{\mathrm{CSA}}(h,a,d) &= \left[1+S_1+S_2-2S_3\right]^{-1}&\label{eq:CSA}\\
	S_1 &= \sum_{n=0}^{\infty}\left(f_{||1}(nd+h,a)^{-1}-1\right)&\nonumber\\
	S_2 &= \sum_{n=1}^{\infty}\left(f_{||1}(nd-h,a)^{-1}-1\right)&\nonumber\\
	S_3 &= \sum_{n=1}^{\infty}\left(f_{||1}(nd,a)^{-1}-1\right)&\nonumber
\end{align}
includes multiple interactions of the perturbations of the pressure and velocity fields
induced by each wall. The same interactions with the colloid are not included.\cite{Lobry96PhysRevB,Lin00PhysRevE}

The lateral diffusion coefficient $D_{||2}$ can be measured from the mean squared displacement (MSD) in one of the orthogonal directions $x$ or $y$. For the $x$-direction and a time interval $\Delta{t}$, the MSD is given by
\begin{equation}
\left<\Delta{}x^2(\Delta{}t)\right> = \left<\frac{1}{N-1}\sum_{i=1}^{N-1 }\left[x(t_i+\Delta{t})-x(t_i)\right]^2 \right> = 2 K_{\alpha}\Delta{}t^\alpha,
\label{eq:diffusion}
\end{equation} 
where $\left<...\right>$ signifies the ensemble average, $N$ is the number of observed positions per trajectory, $K_\alpha$ is a generalized diffusion coefficient and $\alpha$ is the anomalous diffusion exponent~\cite{metzler2000random}. For $\alpha=1$, $K_{\alpha}$ corresponds to the lateral diffusion coefficient $D_{||2}$, however, for $0<\alpha<1$ the behavior becomes sub-diffusive. This situation is best described by a time-scale-depended diffusion coefficient $D_{||2,\alpha}(\Delta{}t) = K_\alpha \Delta{}t^{\alpha-1}$. 

%\begin{align}
%D_{||2}\left(h,a,d\right) &= D_0\,f_{||2}(h,a,d)&\\
%f_{||2}^{LSA}(h,a,d) &= \left[f_{||1}(h,a)^{-1}+f_{||1}(d-h,a)^{-1}-1\right]^{-1}&\\
%f_{||2}^{CSA}(h,a,d) &= \left[1+\sum_{n=0}^{\infty}\left(f_{||1}(nd+h,a)^{-1}-1\right)+&\nonumber\\
%&\sum_{n=1}^{\infty}\left(f_{||1}(nd-h,a)^{-1}-1\right)+&\nonumber\\
%&\sum_{n=1}^{\infty}\left(f_{||1}(nd,a)^{-1}-1\right)\right]^{-1}&
%\end{align}

\subsection{\label{sec:results}Results and Discussion}

\subsubsection{\label{sec:height}Particle height in an asymmetric slit}

\begin{figure}
	\centering
	\includegraphics[width=.60\linewidth]{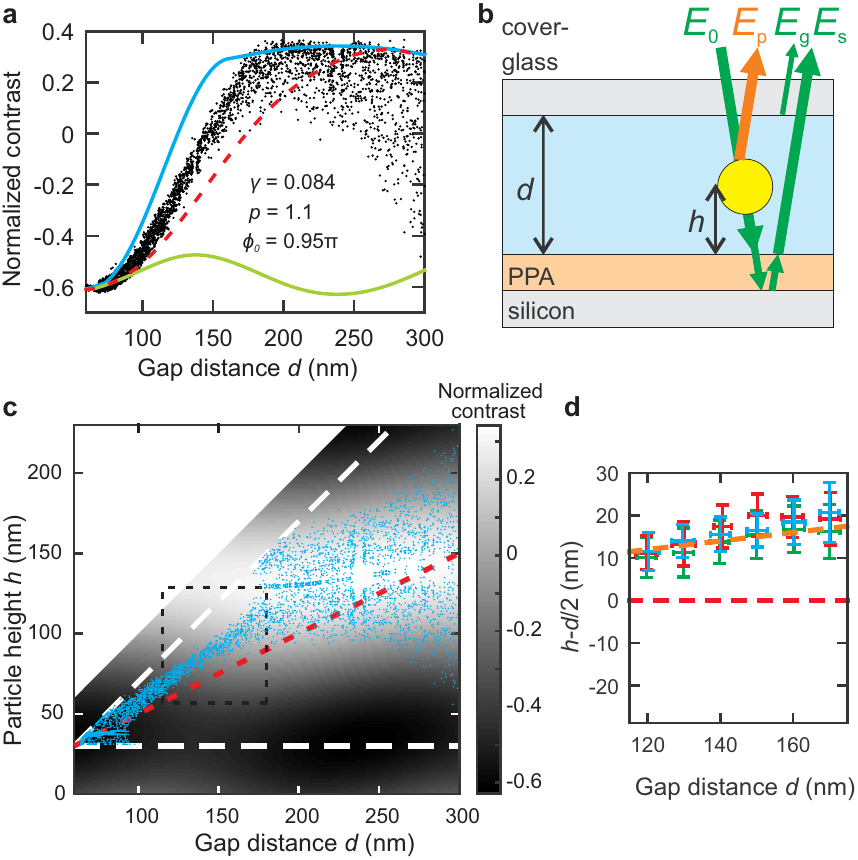}
	\caption{(a) Contrast signal of a nanosphere (dots), a simulated particle in the middle of the gap (dashed red line) and envelopes of simulated maximum (blue line) and minimum (green line) contrast for all possible particle heights. 
	%		(b) SEM image of deposited $60\,$nm Au nanoparticles. The average diameter is $2a = 58.6 \pm 6.2\,$nm. 
	(b) Schematic illustration of the particle at height $h$ in a gap of size $d$. Incoming laser light $E_0$ is scattered by the particle ($E_p$), partially transmitted and reflected at the substrate ($E_s$) and reflected by the cover glass-water surface ($E_g$). 
	(c) Attributed particle heights (blue) are obtained by matching the measured (see panel a) and simulated contrast values (gray-scale image). The confining surfaces and the particle radius restrict the possible $h$ values (dashed white lines). The dashed red line indicates the height values corresponding to the center of the gap. 
	%		A linear fit (red line) to single valued data points (dark blue). The distribution of height variations from the fitted line is shown in the upper inset. 
	(c) Deviation of the observed particle height from the gap center $h-d/2$ for a gap distance range indicated by the dashed black box in panel c for three individually measured particles (blue, red, and green). The error bars indicate the standard deviation of $d$ and $h$. Phenomenologically, the relative particle height follows $h/d \approx 0.61$ (orange dashed line).}
	\label{fig:height}
\end{figure}

According to Eq. (2)-(7), the height $h$ of the particles influences the magnitude of the hindered diffusion. To quantify the effect, we first determine the height for an individually diffusing particle from its contrast. The scatter plot in Fig. \ref{fig:height}\,(a) depicts the experimentally measured and normalized contrast of such a particle for varying gap distance $d$. 

The height of the particle in the gap relates to the contrast that is observed in iSCAT detection. For a fixed gap distance a sinusoidal dependence of the particle contrast with particle height was suggested.\cite{Krishnan10nature} The effect arises from the interference of the light scattered by the particle $E_p$ with the background reflection, that is, the light reflected from the glass $E_g$ and polymer/silicon interface $E_s$, see Fig. \ref{fig:height}\,(b). As discussed in the methods section, the background reflection is also a function of the gap distance, resulting in a more complex relation of the particle contrast with gap distance. In a previous publication we showed how the effective incident angle model describing the background reflection is extended to include the particle refection using three additional parameters to include the light scattered by a nanosphere in the nanofludic gap.\cite{Fringes2016jap}
%Similarly to the determination of the gap distance from the background interference signal, also the height $h$ of a particle - the distance from its center to the polymer surface - can be derived from its intensity. 
%The height $h$ is defined as the vertical distance between the lower polymer-water interface and the center of the particle. 
The first and the second parameter, $p$ and $\phi_0$, describe the amplitude and the accumulated phase of light scattered by the particle and collected by the camera. In addition, at the particle position, the light reflected by the substrate is reduced by a fraction $\gamma$. %The interference of the thus reduced background signal with the light scattered by the particle determines the height dependent particle contrast. 
Due to the interferometric origin, the contrast of the particle is still a periodic function of the particle height with a period of $\omega_L / 2 n_{H_2O} \approx 200\,$nm, where $\omega_L = 532\,$nm is the laser wavelength and $n_{H_2O} = 1.33$ is the refractive index of water. In the experiments, we adjusted the polymer thickness to position the minimum of the particle contrast at tight confinement of $d \approx 70\,$nm, see Fig. \ref{fig:height}\,(a). Consequently, a diffusing particle will probe the entire envelope of the contrast signal if it probes more than 100 nm of the height space above the minimum contrast position. In Fig. \ref{fig:height}\,(a) the black scatter plot indeed does not rise above a particle contrast of $\approx 0.35$ and shows a turnaround at a particle contrast of $\approx -0.6$.
Using the optical model described in detail in Ref. \cite{Fringes2016jap} the parameters $\gamma$, $p$ and $\phi_0$ are iteratively optimized until the envelope predicted by the model (blue and green line in Fig. \ref{fig:height}\,(a)) matches the observed extremal contrast values, considering the finite range of possible particle heights given by the gap distance $d$ and the finite radius $a$ of the particle ($a \leq h \leq d-a$). The procedure ensures that the three parameters can be obtained without the need of additional height calibration using immobilized particles.\cite{Fringes2016jap}
%The fit parameters and the modeled extremal contrast values for all possible particle heights are shown in Fig. \ref{fig:height}\,(a) as solid green and blue lines. 
The red line illustrates the modeled contrast of a particle positioned in the middle of the gap. 
%Three regions can be identified concerning the width of the contrast distribution at a given gap distance. For gap distances between $150<d<300$ the width of the contrast distribution decreases the more the particle is confined by the electrical double layers of the walls. Thus we consider differences in particle height as the dominant origin of contrast variation. For smaller gap distances the width stays constant until there is almost no distribution in the third region for $\lesssim 70\,$nm. While the first distribution and the third look similar for other particles the spread in the second region varies. We interpret the behavior in this region to be dominated by the shape of the particles, which is depicted in the SEM image in Fig. \ref{fig:height} (b). Rotation of asymmetric shaped particles can be the cause of these intensity fluctuations, which abruptly stops once the particle becomes immobilized on one of the confining surfaces. Besides shape, also a size variation of $2a = 58.6 \pm 6.2\,$nm is measured, which agrees with the specifications of the manufacturer (see Sample preparation). Both variations reveal that it is important to measure the contrast and envelopes of individual particles, in order to find their intrinsic parameters.
   
%With the optimal fit parameters of a particle, the contrast can be simulated for any gap distance and particle height. 
The contrast modeled as a function of gap distance $d$ and particle height $h$ is shown as grayscale background in Fig. \ref{fig:height}\,(c). To obtain the height values (blue dots) for a measured contrast we use the simulated values for a given gap distance as a lookup table. The short illumination time of $\lesssim40\,\mu$s is essential to measure almost instantaneous particle heights \cite{Eichmann10langmuir} and to obtain reliable height-distribution data. 
%The particle height is restricted by the physical boundaries given by the confining walls and the radius of the particle (white dashed lines). 
The periodicity of the contrast signal with particle height leads to either one or multiple possible solutions for the particle height. 
%Therefore we only consider the single valued heights within a confinement of $135\,\mathrm{nm}\le d \le 175\,$nm for a linear fit (solid red line). The particle height $h=0.615d$ departs from the middle of the gap and is more repelled by the PPA surface than the glass surface. The asymmetry is also visible by the height distribution in the selected interval as shown in the upper inset in Fig. \ref{fig:height} (c).
%For three independent particles we observe a larger surface potential for the PPA coated surface than for the glass surface. The increased surface potential leads to more repulsion of the particles visualized by the relative height depicted in the inset. 
In the single-value range of $115\,\mathrm{nm}\le d \le 175\,$nm we determined the averaged deviation $h-d/2$ of the particle height from the gap center (see Fig. \ref{fig:height}\,(d)).

Physically, the average height of the negatively charged particles is determined by the relative repulsion of the particles from the like charged confining surfaces. A height above the center of the gap indicates a higher charge on the polymer surface, which does not contain sites that could dissociate. However, it is known that hydrophobic surfaces often attain a negative charge in contact with water, most likely due to the preferential absorption of oxianions.\cite{tian2009structure}

\subsubsection{\label{sec:diffusion}Confined lateral diffusion of nanospheres}
To measure the lateral diffusion of nanoparticles as a function of gap distance, we exploit the high mechanical stability and tunability of the nanofluidic confinement apparatus.
%The approximations about hydrodynamic hindrance (see Theory) are made for flat surfaces and distinct wall separations. 
We vary the gap distance for different measurements and then use the feedback-control loop to keep it constant (see Fig. \ref{fig:performance}\,(e)) while acquiring frames for $15\,$s. For gap distances $d\gtrsim200\,$nm, on average $23\pm 5$ particles per frame are detected, whereas for higher confinements with $d\lesssim200\,$nm only $8\pm3$ particles are detected. The high frame rate (800 FPS) nevertheless provides a sampling of 60,000 up to 300,000 particle positions for each measurement.

\begin{figure}%
	\centering
	\includegraphics[width=.62\linewidth]{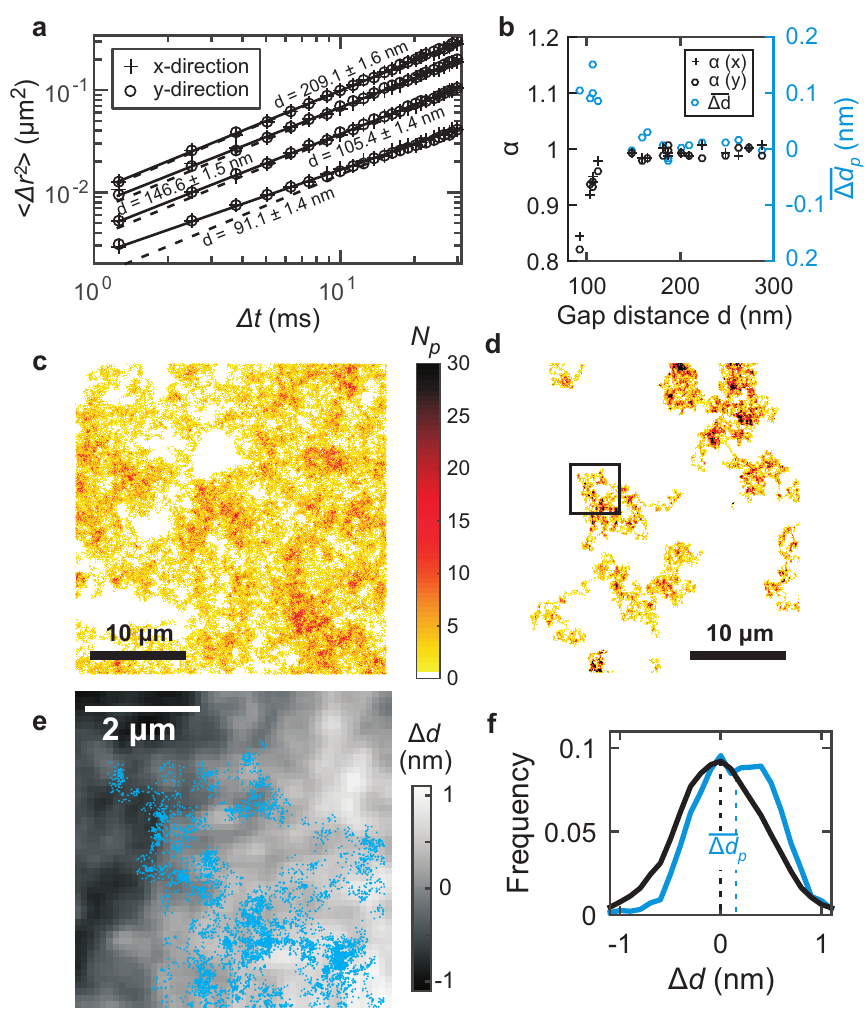}
	\caption{(a) Measurements of the lateral MSD in the $x$- and $y$-direction at four different gap distances $d$. The solid and the dashed lines indicate fits of Eq. (\ref{eq:diffusion}) to the data for anomalous and normal diffusion, respectively. (b) The parameter $\alpha$ indicates the degree of normal diffusion $\alpha=1$ and sub-diffusion $\alpha<1$ (black symbols). An average value of the $\overline{\Delta{}d_p} > 0$ (see panel f) indicates that particles avoid narrow gap regions (blue circles). (c,d) Number of detected particles $N_p$ in a $100\,$nm grid during a $15\,$s measurement at an average gap distance of (c) $d=210.0\pm1.0\,$nm and (d) $d=105.9\pm1.0\,$nm. (e) Gap distance modulation $\Delta{}d$ and detected particle positions (blue dots) for the area indicated by the black box in panel d. (f) Histograms of the gap distance modulation (black line) and for the locations of the gap distance modulation sampled by the particles (blue line).}
	\label{fig:diffusion}
\end{figure}

For each gap distance $d$, we obtain the one-dimensional (1D) time and ensemble averaged MSD for a range of time steps $\Delta t$ from $1.25 \leq \Delta t \leq 31.25\,$ms, see Fig. \ref{fig:diffusion}\,(a). A strong decrease of the diffusivity with decreasing gap distance is apparent. %The MSD is measured for the independent x- and y-dimension. 
Fits of Eq. (\ref{eq:diffusion}) to the MSD in the $x$- and $y$-directions are given as solid lines. The dashed lines indicate fits to the data for normal diffusion ($\alpha = 1$). The fit parameter $\alpha$ indicating sub-diffusion for $\alpha<1$ is shown in Fig. \ref{fig:diffusion}\,(b). At a confinement $d<120\,$nm, a scale-dependent diffusion coefficient is observed, see also the increasing deviation of the dashed and solid lines in Fig. \ref{fig:diffusion}\,(a). This effect has been attributed to the presence of lateral obstacles preventing a free diffusion of the particles.\cite{Volpe14brownian} In our case however, these obstacles are either induced by local charge inhomogeneities or by the roughness of the confining walls.

We use a simple picture to assess this hypothesis. In the so called linear superposition approximation the interaction energy $U(h)$ of a charged spherical particle at a distance $h$ to a charged plane is given by:\cite{Fringes2016jap,adamczyk1996role}

\begin{equation}
\label{phitotal}
U(h) = 4 \pi \epsilon \epsilon_0 a \psi_{P,eff} \psi_{S,eff} e^{-\kappa (h-a)},
\end{equation}

where $\kappa^{-1}$ is the Debye length, $\epsilon$ is the dielectric constant of the medium, $\epsilon_0$ is the vacuum permittivity, $a$ is the radius, and $\psi_{P,eff}$ and $\psi_{S,eff}$ are the effective surface potentials of plane and sphere, respectively. In this linear approximation the overall interaction energy of a sphere between two walls is obtained by the sum of the interaction energies to each wall. Assuming a surface potential of the sphere of $-58 mV$ (see methods) and a surface potential of the walls of $-67\,$mV as determined in our previous experiments \cite{Fringes2016jap}, we obtain a change in interaction energy of $\approx 0.8\,k_B T$ for a gap distance of $120\,nm$ and a gap distance modulation of $1\,$nm. The simple model corroborates the interpretation that the observed RMS roughness of the glass of 0.4 nm provides significant energy barriers for diffusion. We note, however, that the same effect could be induced by a charge modulation of the surface potential (or correspondingly the surface charge) by $\approx 5 \%$.

To further investigate the origin of the obstacles we analyzed the time-averaged lateral particle distribution and its correlation to the measured locally resolved gap distance variation $\Delta d$ (see Fig. \ref{fig:performance}d). 
To obtain a measure for the particle distribution, we divide the field of view into a $100\,$nm grid and count the total number of particles visiting each grid section for all frames. The resulting number of detected particles is visualized as "heatmaps" in Fig. \ref{fig:diffusion}\,(c,d) for an average gap distance of (c) $d=209.1\pm1.0\,$nm and (d) $d=105.4\pm1.0\,$nm. The particles are quasi uniformly distributed over the entire field of view for the larger separation and are more localized in the narrower slit. 
%Therefore we like to investigate if there is a gap distance dependent correlation between the localization and the local gap distance modulation. 

%optical path difference in the nanofludic gap at diffraction-limited resolution. The local variation of the optical path difference is given by the roughnesses of the silicon, the polymer, and the glass. Because the glass has the highest surface roughness, we infer that the roughness of the optical path difference image is also dominated by the glass roughness. Assuming that all the roughness is provided by the glass, we can use the optical model to calculate a map of gap distances from the optical path differences. 

In order to correlate the detected particle trajectories with the gap distance modulation $\Delta d$, see Fig. \ref{fig:performance}\,(d),  we have to compensate for the tilt in the gap distance map. We divide the map into squares of $5\times5\,\mu$m$^2$ size, roughly corresponding to the 1D diffusion length during the measurement of $r_{\mathrm{diff}} \approx 5\,\mu\mathrm{m}$,  and correct for the offset in local gap distance modulation $\Delta d$ for each square. For example, Fig. \ref{fig:diffusion}\,(e) shows $\Delta{}d$ and the positions of a single diffusing particle (blue dots) for the square given by the box in Fig. \ref{fig:diffusion}\,(d). According to this trace the particle samples certain locations of the $\Delta d$ map and we term the range of sampled values $\Delta d_p$.
%the average gap distance modulation within $3\,\mu$m is subtracted from the calculated gap distance modulation (see Fig. \ref{fig:performance}(d)). 
The average histograms of $\Delta d$ and $\Delta d_p$ for all squares are shown in Fig. \ref{fig:diffusion}\,(f) as black and blue lines, respectively. Clearly, the particles prefer to be located at a position having a larger gap distance as apparent by the shift of the $\Delta d_p$ histogram to more positive $\Delta d$ values. To obtain a qualitative measure of the strength of this effect, we determined the distance of the center of mass of the two histograms $\overline{\Delta{}d_p}$ for all measured gap distances, see Fig. \ref{fig:diffusion}\,(f). The result is given in Fig. \ref{fig:diffusion}\,(b) by the blue circles. For gap distances below $d = 120$\,nm, a significant shift of the particle position into high-gap-distance positions is apparent. This behavior is qualitatively similar to the onset of sub-diffusion measured for the MSD. Therefore we conclude that the sub-diffusion is indeed caused by the fact that the particles start to avoid regions with narrower gap distances. 
%At first glance this seems surprising because the particle-wall distance at the onset of sub-diffusion is on the order of 30 nm and thus much bigger than the surface roughness; however, one has to keep in mind that the particle-surface interactions increase exponentially with decreasing particle-wall distance.

\begin{figure}%
	\centering
	\includegraphics[width=.7\linewidth]{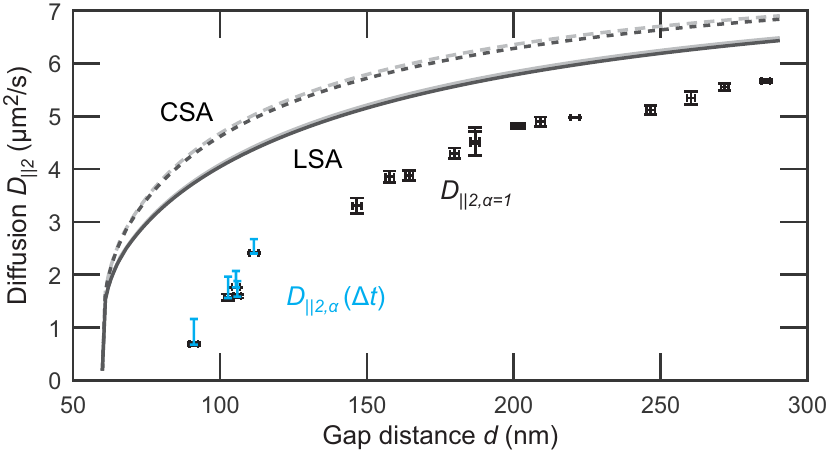}
	\caption{Lateral diffusion coefficients for varying gap distance in a nanofludic slit. The error of the normal diffusion coefficients $D_{||2}$ (black scatter plot) and $d$ are determined by the difference of $D_{||2}$ in x and y direction and the standard deviation of $d$, respectively. The range of time dependent diffusion coefficients $D_{||2,\alpha}(\Delta{}t)$ for $1.25\,$ms$\,<\Delta{}t<\,31.25\,$ms is plotted for $d < 120 \,$nm (blue bars). For greater gap distances the range of $D_{||2,\alpha}(\Delta{}t)$ is less than 8\% of the normal diffusion coefficient. Theoretically predicted diffusion coefficients by LSA (solid lines, Eq. \ref{eq:LSA}) and CSA (dashed lines, Eq. \ref{eq:CSA}) are shown for an average particle height at $h=0.5d$ (gray) and $h=0.61d$ (black).}
		%		The green lines considers an effective particle diameter of $a_{eff}=a+\delta$ and an effective gap distance $d_{eff}=d-2\delta$. The meaning of line style and brightness is the same as in gray.		
	\label{fig:diffusion2}
\end{figure}

Now we turn to the central result, the gap-distance-dependent lateral diffusion coefficient $D_{||2}(d)$, which is depicted in Fig. \ref{fig:diffusion2}. The black scatter plot indicates the values for normal diffusion $D_{||2,\alpha=1}(d)$ corresponding to the dashed lines in Fig. \ref{fig:diffusion}\,(a). For $d < 120\,$nm sub-diffusion is significant and a single diffusion coefficient is not sufficient to describe the process, see Eq. (\ref{eq:diffusion}). Instead, the diffusion coefficient $D_{||2,\alpha}(d,\Delta{t})$ becomes dependent on the time interval $\Delta{t}$. The range for $D_{||2,\alpha}(d,\Delta{t})$ for $1.25\,$ms$\,<\Delta{}t<\,31.25\,$ms is indicated for $d < 120\,$nm by the blue bars.

For comparison, the predicted diffusion coefficients accounting for hydrodynamic hindrance from two walls are shown for the LSA [Eq. (\ref{eq:LSA}) (solid lines)] and CSA [Eq. (\ref{eq:CSA}) (dashed lines)]. Both approximations were calculated for a particle diffusing at a measured height $h=0.61d$ (black) and in the middle of the slit $h=0.5d$ (gray). The asymmetric height leads to merely $1.5\%$ lower diffusion coefficients and cannot explain the $20-50\,\%$ lower diffusivity measured. We also exclude that the localization due to surface roughness is the predominant factor for this reduction, because pronounced sub-diffusion is only observed for gap distances of $d<120\,$nm. 

In bulk, the electroviscous effect is attributed to the surface charge of the particles and leads to an increased effective viscosity and thus to a reduction in particle diffusion.\cite{conway1960rheology} A similar mechanism should also play a role in a nanofluidic system, in particular when a particle is close to a charged wall. Whereas diffusion measurements for uncharged particles\cite{Lin00PhysRevE} and for particles in electrolyte with higher ionic concentration\cite{Eichmann10langmuir} are in agreement with predictions that consider only a hydrodynamically hindered drag. There is considerable evidence of an increased drag of charged particles near charged walls in a weak electrolyte.\cite{Carbajal-Tinoco97PhysRevE,Eichmann08lgm}  In a similar experimental configuration Eichmann \textit{et al.} \cite{Eichmann08lgm} measured a $\approx30$\,\% ($\approx55$\,\%) lower lateral diffusion coefficient for 60\,nm (100\,nm) gold nanospheres with a relative radius of $\kappa a \approx 0.9$ ($\kappa a \approx 2.1$) and a relative glass-particle distance of $\kappa h-\kappa a \approx 4.5$ ($\kappa h-\kappa a \approx 3.6$). These values are in agreement with the $\approx45$\,\% lower diffusion we measure for $\kappa a \approx 3.4$ and $\kappa h-\kappa a \approx 4$.

\section{Conclusion}
We have developed a new versatile setup for investigating the behavior of nano-objects in a tunable confinement between two surfaces. The interferometric detection setup allows us not only to detect the nano-objects with high sensitivity, but also to determine the 3D particle position and the wall separation \emph{in situ} with nanometer spatial and millisecond temporal precision. Furthermore, a diffraction limited resolved map of the sub-nanometer-resolved gap distance can be obtained. We use the tool to measure the height and diffusion of $60\,$nm gold spheres as a function of absolute gap distance between a glass and a polymer surface. We find that the particles localize more closely to the glass interface indicating a higher charge of the polymer surface. Sub-diffusion becomes significant at gap distances below $d = 120\,$nm. We demonstrate that this scale dependent diffusion is correlated to particle trajectories that avoid regions of narrow gap distances caused by the surface roughness of the confining surfaces. The measured lateral diffusion coefficients are $20 - 50\,\%$ lower than predicted by purely hydro-dynamical hindrance, also when taking their asymmetric position in the gap into account. Similarly, the observed scale dependent diffusion cannot account for the effect because it is only significant for small gap distances. We conclude that electro-viscous effects are the main cause for the observed reduction in diffusivity. Our measurements provide a detailed information on the gap-distance-dependent particle diffusion, which may form the basis for testing theories describing the electro-viscous effect. In general, the results shown here demonstrate the versatility of the tool which allows one to measure nano-particle behavior as a function of confinement in remarkable detail. 

%The most likely the electroviscous effect leads to the greater than expected hindrance of the particle motion. Nevertheless an influence on diffusion is also visible from the sub-nanometer surface roughness. Therefore deliberately shaping the topography of the PPA surface with a nanometer accuracy by means of lithography \cite{Pires10sci} will allow us to strongly influence the diffusion and movement of the particles in a target way, enabling sorting or assembly of particles with high accuracy.

\begin{acknowledgments}
The authors thank U. Drechsler, M. Sousa, and S. Reidt for technical support, C. Bolliger for proof-reading, and U. Duerig and M. Krishnan (University of Zurich) for fruitful discussions. Funding has been provided by the European Research Council StG no. 307079.
\end{acknowledgments}

\appendix

\nocite{*}
\bibliography{nca}% Produces the bibliography via BibTeX.

\end{document}